\documentclass[10pt]{article}

\usepackage[english]{babel}

\usepackage{authblk}

\usepackage[utf8]{inputenc}
\usepackage{breakcites}
\usepackage{graphicx}

\usepackage{amsmath,amssymb}

\usepackage{nameref}
\usepackage[breaklinks=true]{hyperref}

\usepackage{xcolor} % Required for highlighting changes
\usepackage{listings}
\newcommand{\code}[1]{\texttt{#1}}

\newif\ifnotes\notestrue
\notesfalse
\def\boxnote#1#2{\ifnotes\fbox{\footnote{\ }}\ \footnotetext{ From #1: #2}\fi}

\def\Edwin#1{\boxnote{\color{blue}Edwin}{#1}}

\begin{document}

\title{PyGOM --- A Python Package for Simplifying Modelling with Systems
of Ordinary Differential Equations}
\author[1]{Edwin Tye \thanks{Edwin.Tye@phe.gov.uk}}
\author[2]{Tom Finnie \thanks{Thomas.Finnie@phe.gov.uk}}
\author[3]{Ian Hall}
\author[4]{Steve Leach}

\affil[1,2,3,4]{Emergency Response Department Science \& Technology, Public Health England}
%\affil[2]{Emergency Response Department Science \& Technology, Public Health England}
%\affil[3]{Emergency Response Department Science \& Technology, Public Health England}
%\affil[4]{Emergency Response Department Science \& Technology, Public Health England}

%\author{
%    Edwin Tye\\
%    \texttt{Edwin.Tye@phe.gov.uk}
%    \and
%    Tom Finnie \\
%    \texttt{Thomas.Finnie@phe.gov.uk}
%    \and
%    Ian Hall \\
%    \and
%    Steve Leach
%}

\maketitle

\section*{Abstract}
Ordinary Differential Equations (ODE) are used throughout science where the
capture of rates of change in states is sought. While both pieces of commercial
and open software exist to study such systems, their efficient and accurate
usage frequently requires deep understanding of mathematics and programming. The
package we present here, PyGOM, seeks to remove these obstacles for models based
on ODE systems. We provide a simple interface for the construction of such
systems backed by a comprehensive and easy to use tool--box. This tool--box
implements functions to easily perform common operations for ODE systems such as
solving, parameter estimation, and stochastic simulation. The package source is
freely available and organized in a way that permits easy extension. With both
the algebraic and numeric calculations performed automatically (but still
accessible), the end user is freed to focus on model development.

% \end{section*} Abstract

% [Introduction]
\section*{Introduction}
In this paper we introduce a Python package, PyGOM (Python Generic ODE
Model, \code{pygom} in code); a toolbox for modeling with Ordinary
Differential Equations (ODEs). This package enables the user to define models
made from systems of ODEs in a mathematically intuitive manner that makes
interactive investigation simple.  Once defined, such a system may be
solved and used to provide realizations with either parameter or jump process
stochasticity. Parameters, complete with calculations of confidence intervals,
may be easily estimated from data. This package is designed to make the
construction, parametrization, manipulation, visualization and solving of ODE
based models as uncomplicated as possible for the end user.

PyGOM's was initially created so that during disease outbreaks the models
presented in the literature may be rapidly and rigorously validated in the
absence of source code. Disease outbreaks of international concern such as the
West--African Ebola epidemic \cite{who_ebola_response_team_ebola_2014,
who_ebola_response_team_west_2015}, Middle--East Respiratory
Syndrome \cite{cauchemez_middle_2014} or the 2009 Influenza A H1N1 pandemic
\cite{fraser_pandemic_2009} cause a great many papers to be produced and the
political decision making process demands a speedy and robust scientific
analysis of these so that mitigation and emergency response operations may be
performed. PyGOM has grown far beyond this genesis to become a general toolkit
for working with ODE systems in the many places they occur.

Although PyGOM has its roots deep in epidemiology modeling, we
recognize that the application of ODE is vast, and different communities have
developed their own way of distributing existing models.  Notably
\href{http://sbml.org/Main_Page}{SBML} \cite{hucka_systems_2003} and
\href{https://www.cellml.org/}{CellML} \cite{miller_overview_2010} have
significant followings and translation between the two is possible
\cite{smith_sbml_2014}. PyGOM has the ability to read and write simple SBML
documents and there are plans to extend the package to accommodate the full set
of features.

%     - What are ODES
ODEs are differential equations with a function or functions containing a single
independent variable and its derivatives. The term ``ordinary'' is used to
distinguish these equations from partial differential equations where there can
be more than one independent variable. ODEs can be written in the general form
of
\begin{equation*}%\label{eqn:ode}
    \frac{dx}{dt} = f(x)%\notag.
\end{equation*}

ODEs are used across all scientific disciplines as they are a natural
way to describe change and rates of change of quantities in a precise and
concise mathematical form. As such ODEs are a well studied area and we refer
interested readers to \cite{Lambert1973,Stroud2004,Robinson2004} for
introductory textbooks or \cite{Coddington1984,Hairer1993,Hairer1996} for the
more advanced topics. 

%     - What is the motivation for the package
Differential equations may be coupled into systems. Such systems of ODEs are
used extensively across all numerical sciences to model physical systems and
processes. For example, most compartmental models may be formulated as systems
of ODEs. Solving these ODEs and ODE systems can be broadly split into Initial 
Value Problems (IVP) and Boundary Value Problems (BVP). As solving a BVP
can be viewed as parameter estimation in IVP, our focus is solely on IVP within
this package.

Modeling using ODEs is a relatively mature area, resulting in the existence of
commercial software such as APMonitor \cite{Hedengren2014}, MATLAB and many
others. However, restrictive licensing and cost considerations limit their
accessibility and inhibit their use in the wider ecosystem of open--source
analytic tools. This is particularly acute in High Performance Computing
environments where per--CPU cost becomes rapidly restrictive. Indeed,
even trial--to--paid toolboxes like PotterWheel \cite{Maiwald2008} or free
ones such as Systems Biology Toolbox \cite{Schmidt2006} still require MATLAB.
Other alternatives such as Sundials \cite{Hindmarsh2005} provides a
\verb|C| interface, and are exceptional in terms of computation speed but
are not friendly when models are being rapidly developed and tested.

Performance of our program is platform dependent as the type of
compilation that can be achieved for the functions will differ between machines. 
The time required to perform one function evaluation is typically reduced to a
quarter of what is required of pure Python code such as
\href{http://pysces.sourceforge.net/}{PySCeS}.  A function evaluation here can
be say the $f(x)$, $\nabla_{x} f(x)$ or other related information.
% Evidently, software like \href{http://copasi.org/}{COPASI} which
% was built on \verb|C++| and access via Python bindings have similar speed.

With PyGOM we sought to address these limitations by producing a complete system
that allows the rapid design, prototyping and use of such ODE models. We
harness the many capabilities of Python and its packages --- the fast
prototyping ability of a dynamic programming language, manipulation of algebraic
expressions, the ability to compile these expressions to static programming
languages during run--time for performance, running model
realizations in parallel and good visualization tools --- while keeping the
interface simple and intuitive. The software itself is accessible to all under 
an Open Source license, freeing it to be used without restriction on desktops,
cloud systems and even in High Performance Computing environments.  

% \end{section} % Introduction

% Goal of the package and brief comparison (difference) to other packages
\section*{Overview}
The amount of existing software focused on ODE modeling is
vast and ever expanding.  Nearly all are created with a particular focus,
tailored to the creators' field of expertise.  Given the fast moving pace of the
software development world, to make a sound comparison with all existing
software is impossible.  Instead, we quickly walk through the key feature set of
PyGOM here with further exposition in later sections.

With the initial motivation stemming from evaluation of models during disease
outbreaks, the design and feature set is catered towards epidemiology.
More concretely, we were faced with tasks such as performing sanity checks
on models and calculating simple information such as basic reproduction
number (R0).  This can be challenging given different ways of describing
systems of equations.

% Defining the model through a stict class structure as
% shown previously (less so than SBML) may seen more cumbersome when compared
% to simply writing it out in MATLAB, or even Python.

PyGOM has the capability to decompose the model from the ODE form into
individual transitions which can then be used to perform stochastic simulation
or model verification.  Various analyses can then be performed on the
transitions; in terms of algebraic manipulation or numerical evaluation
if the parameters are known.

In the event that the parameters of a model are not known, estimating
them from data is also possible.  For convenience, PyGOM has the capability to
read EpiJSON data \cite{finnie_epijson:_2016} directly, providing a more
robust data interchange than free text formats.

Reporting the point estimate of parameters with epidemiological
meaning such as the incubation period can be misleading.  Multiple ways of
obtaining confidence intervals (CI) on parameter estimates are provided in the
package.  They have been designed to be easy to use such that a CI can
be routinely reported. We demonstrate the functions later in this paper to
show the work flow; from parameter estimation to generating
the corresponding confidence intervals using convenient artificial data. 
Further examples and details are available in the package documentation.

Using Python as a development platform permits the end user to
develop a model dynamically.  In particular it eases the construction of
multi--type models such as the SIS model (later section).  This is because we
can generate the set of states using list comprehension
\begin{lstlisting}[language=Python]
    >>> types = ['v','h']
    >>> state = [x+'_'+i for x in ['I','S'] for i in types]
['S_v', 'S_h', 'I_v', 'I_h']
\end{lstlisting}
and changing types to say country name or age group is trivial. Vector
notation may also be used in PyGOM.

%\green{Compare features against other software, waiting for our development to
%progress.}

% \end{section} % Overview

% [Basic example]
\section*{Basic usage}
\begin{figure}[bt]
\centering
\includegraphics[width=5in]{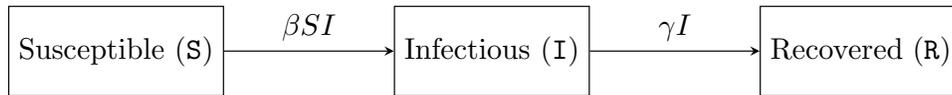}
%% \hspace*{10em}
\caption{A block diagram of the very simple SIR disease model
used in examples throughout this paper. This model contains
three states: susceptible (\code{S}), infectious (\code{I}) and
recovered (\code{R}) with transition between these states
controlled by two parameters: the infection rate ($\beta$) and
the recovery rate ($\gamma$).}\label{fig:sir_block_diagram}
\end{figure}

As an introduction to PyGOM we use the standard SIR compartmental model
\cite{Brauer2008}. A block diagram of this model is presented in Fig.
\ref{fig:sir_block_diagram}.

This consists of three disease states: susceptible ($S$), infectious ($I$) and
Recovered ($R$) and three parameters: infection rate ($\beta$), recovery rate
($\gamma$) and total population ($N$). The total population is usually omitted
from the SIR model definition, but it is convenient to include it here for
demonstration purposes.  The model is defined through the following two
transitions
\begin{align*}
  S \to I &= \beta S I/N \\
  I \to R &= \gamma I.
\end{align*}
For simplicity we have not used birth or death processes here but the
inclusion of such mechanisms in a model is possible and they will be
introduced later. Below we define this system from first principles. However,
we have provided a set of commonly used models in PyGOM's
\code{common\_models} module and within this module a predefined version
of the SIR model may be found. Greater detail on these models within the
module has been provided in the supplementary material.

\subsection*{Model construction}
To construct this model we begin by importing the PyGOM package and defining
these transitions with the first in the more expressive form:
\begin{lstlisting}[language=Python] % block 1
    >>> from pygom import Transition, TransitionType
    >>> sir_t1 = Transition(origin='S',  
                            destination='I',
                            equation='beta*S*I/N',
                            transition_type=TransitionType.T) 
    >>> sir_t2 = Transition('I', 'gamma*I', 'T', 'R')
\end{lstlisting}
We now need to define the states and parameters
in this model.
These are simply defined as lists.
\begin{lstlisting}[language=Python]
    >>> states = ['S', 'I', 'R']
    >>> params = ['beta', 'gamma', 'N']
\end{lstlisting}
No further information is required to define the SIR model. We may
now initialize and verify the model. The initialized class will convert the 
equations provided in the \code{Transition} objects into algebraic form using
the \code{sympy}\cite{sympy} package. Our classes automatically translate
the equations from symbolic to numerical form by run--time compilation. 
Significant differences in performance may be observed depending on the setup of
the machine on which PyGOM is being used. In particular, the availability of
\verb|FORTRAN| and \verb|C| compilers.
\begin{lstlisting}[language=Python]
    >>> from pygom import DeterministicOde
    >>> model = DeterministicOde(states,
                                 params,
                                 transition=[sir_t1, sir_t2])
    >>> model.get_ode_eqn()
Matrix([
[       -I*S*beta/N],
[I*(S*beta/N-gamma)],
[         I*gamma]])
\end{lstlisting}
The equations returned by \code{get\_ode\_eqn()} correspond to the states and
their order as defined in \code{states}. In addition, to show the output
in English we provide \code{print\_ode()}. By default this displays the system 
in symbolic form but by changing the input argument of \code{latex\_output} to
\code{True}, the corresponding equations in latex form will be shown instead. 
This is to eliminate the need to type out the equations again at a later date. 
Further information, such as the Jacobian and gradient for the system of ODEs,
are provided by PyGOM through the model object and may be obtained using the
\code{get\_jacobian\_eqn()} and \code{get\_grad\_eqn()} methods respectively.
%non deterministic
\begin{lstlisting}[language=Python]
    >>> model.get_jacobian_eqn()
Matrix([
[-I*beta/N,         -S*beta/N, 0],
[ I*beta/N, -gamma + S*beta/N, 0],
[        0,             gamma, 0]])

    >>> model.get_grad_eqn()
Matrix([
[-I*S/N,  0,  I*S*beta/N**2],
[ I*S/N, -I, -I*S*beta/N**2],
[     0,  I,              0]])
\end{lstlisting}

Alternatively, we can also define the SIR model via a set of explicit
ODEs.  We omit the details here as the setup is similar to the vector--host
model shown later.

\begin{lstlisting}[language=Python]
    >>> ode1 = Transition(origin='S',
                          equation='-beta*S*I/N',
                          transition_type=TransitionType.ODE)
    >>> ode2 = Transition('I', 'beta*S*I/N-gamma*I',
                          TransitionType.ODE)
    >>> ode3 = Transition('R', 'gamma*I', 'ODE')
    >>> model = DeterministicOde(states,
                                 params,
                                 ode=[ode1, ode2, ode3])
\end{lstlisting}

%% \end{subsection}{Model Construction}

%     - solve
\subsection*{Solving the model}
The most common use of an ODE is to generate a solution for an IVP. That
is, given an initial time point $t_{0}$ and corresponding observation
$x(t_{0})$, a set of solutions is found for some time $\mathbf{t} =
\left[t_{1},\ldots,t_{n}\right]$.
An analytical solution is attainable when $f(x)$ is linear, otherwise a
numerical integration is required. We refer to such solution as a
\emph{deterministic}. To test a system's linearity we simply ask the ODE
object
\begin{lstlisting}[language=Python]
    >>> model.linear_ode()
False
\end{lstlisting}
That this is False comes as no surprise as we know the SIR model is
non--linear.
% To solve non--linear systems we have to use numerical integration which we
% refer to as a deterministic solution. 
The following example is taken from \cite{Brauer2008}. We define the values of
the parameters and the initial conditions as preparation for the evaluation of
the IVP. It is important to note at this point that the numeric values of the
states need to be set in the correct order against the list of states, which is
the same as defined when the model was created.
\begin{lstlisting}[language=Python]
    >>> N = 7781984.0
    >>> init_state = [0.065*N,123*(5.0/30.0),0.0]
    >>> param_eval = [('beta', 3.6), ('gamma', 0.2), ('N',N)]
\end{lstlisting}
We are usually interested in how the states within the model change
over time. First we used the Python package \code{numpy}'s \code{linspace}
function to create an evenly spaced time vector between $t=0$ and $t=150$. We
then inform the model object of the initial conditions and parameter values, and
finally solve the problem using the model's integrate function
\begin{lstlisting}[language=Python]
    >>> import numpy as np
    >>> t = np.linspace(0, 150, 100)
    >>> model.initial_values = (init_state, t[0])
    >>> model.parameters = param_eval
    >>> solution = model.integrate(t[1:])
\end{lstlisting}

%% \end{subsection}{Solving the model}

\subsection*{Alternative integrators}
Internally PyGOM makes use of the integrators provided by the SciPy package and
provides a simple interface to this functionality. As SciPy make use
of odepack, the \emph{de facto} standard, the speed of the integration is only dependent
on each function call. However the methods chosen by PyGOM's internal integrator
may not be suitable for all possible ODE systems.\Edwin{or analysis for that
matter, i.e. we may want to use LSODAR to find the root of the Jacobian} By
using the exposed methods of the model object, namely \code{ode} and
\code{Jacobian}, we allow end users to use any integration algorithm of their
choice.  The two aforementioned methods take two input arguments $(x,t)$ the
state and time respectively.  All the available methods exported from the model
also have a complement, the same function name with a `\textbf{T}' appended to
the end which take the same arguments but in the reverse order. As an example,
to perform the same analysis as the internal integrate function using SciPy's
standard numerical integrator, \code{odeint} we would do the following
\begin{lstlisting}[language=Python]
   >>> from scipy.integrate import odeint
   >>> sol_ext = odeint(model.ode, init_state, t[1:])
\end{lstlisting}

%     - plot
\subsection*{Plotting a model}
\begin{figure}[!h]
\includegraphics[width=5in]{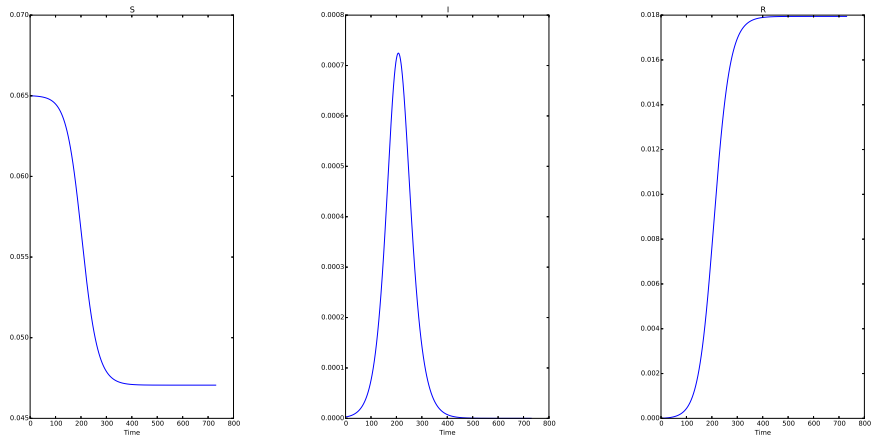} % {Fig2}
\caption{Solution of a simple SIR model}\label{fig:sir_plot}
\end{figure}

To simplify visualization of an initialized ODE system we supply the
\code{plot()} function. This takes advantage of \code{matplotlib} to display the
results in a compact manner
\begin{lstlisting}[language=Python]
    >>> model.plot()
\end{lstlisting}
If more control of plotting is required then the values of the states may be
taken from the solution object to produce graphs such as Fig.
\ref{fig:sir_plot}. This figure was produced using the same method as PyGOM's
internal plot function and only differs from the result of PyGOM's \code{plot()}
in the naming of the axes. However, as the values are available, any graphing
program could have been used.

%% \end{section}{Base Usage}

\section*{Epidemiology focused features}
PyGOM can decompose a set of ODEs into individual transitions
between states and birth/death processes.  Consider a simple vector--host SIS
model \cite{Brauer2008}
\begin{align*}
S_{h}^{\prime} &= \lambda_h + \mu_h S_h-\beta_h S_h I_v + \gamma I_h \\
S_{v}^{\prime} &= \lambda_v + \mu_v S_v-\beta_v S_v I_h \\
I_h^{\prime} &= \beta_h S_h I_v - (\mu_h + \gamma) I_h) \\
I_v^{\prime} &= \beta_v S_v I_h - \mu_vI_v,
\end{align*}
under Lagrange's notation.  This can be entered into PyGOM as
\begin{lstlisting}[language=Python]
    >>> from pygom import SimulateOde, Transition as T
    >>> state = ['S_v', 'S_h', 'I_v', 'I_h']
    >>> param = ['beta_v', 'beta_h', 'mu_v', 'mu_h',
                     'lambda_v', 'lambda_h', 'gamma']
    >>> t1 = T('S_h', 'lambda_h-mu_h*S_h-beta_h*S_h*I_v+gamma*I_h') 
    >>> t2 = T('S_v', 'lambda_v-mu_v*S_v-beta_v*S_v*I_h') 
    >>> t3 = T('I_h', 'beta_h*S_h*I_v-(mu_h+gamma)*I_h')
    >>> t4 = T('I_v', 'beta_v*S_v*I_h-mu_v*I_v')
    >>> ode = SimulateOde(state, param, ode=[t1,t2,t3,t4])
\end{lstlisting}
where the last line initializes the model.  Some of the standard operations
such as simulating the ODE can be performed and will be discussed later.

We show how an R0 can be obtained by calling the corresponding
methods, given the disease states as per the second line below.
\begin{lstlisting}[language=Python]
    >>> from pygom.model.epi_analysis import R0
    >>> ode = ode.get_unrolled_obj()
    >>> R0(ode, ['I_v','I_h'])
sqrt(beta_h*beta_v*lambda_h*lambda_v/(mu_h*(gamma+mu_h)))/Abs(mu_v)
\end{lstlisting}
The R0 value above has already made the substitution for the states using
the disease free equilibrium (DFE).  Algebraic expression for the DFE can be
obtained on its own, and the output would have been numerical instead of
symbolic if the parameter values were available. Note that the \code{ode}
object has been replaced in the first line and is now composed of transitions
between states and birth/death processes.  We can visualize the model or
perform manipulation (such as deleting a death process) with this new object.

\section*{In depth usage}
%     [Transitions]
\subsection*{Transitions and the transition object}
Fundamental to setting up a model is to correctly define the set of ODEs that
are to be built into the system. Within PyGOM these are defined using the
\code{Transition} object defined in the \code{transition} module. The
construction of such an object takes a number arguments but the four
most important ones are:
\begin{enumerate}
  \item The origin state (\code{origin})
  \item An equation, as a string, that defines the process (\code{equation})
  \item The type of transition (\code{transition\_type})
  \item The destination state (\code{destination})
\end{enumerate}
When constructing a \code{Transition} object, two
arguments are required with two optional arguments: \code{transition\_type},
\code{destination}, defaulting to an 'ode' and None. While we have only showed a
transition between two states, both the origin and destination can accommodate
multiple states to represent transitions like $A + A \rightarrow B + C$. In the
example above we showed that the SIR model could be constructed using either the
equations of the transition between states using a \code{Transition} with type
T, or by defining the ODEs that control the states using a transition of type
ODE. Two further types of transitions are possible, birth and death processes,
which are types B and D respectively. These add to or remove from a state
without a source or destination state.

Defining the model through a class structure is no more difficult than
say MATLAB or Python in their plain equation form.  Although some of the
code samples shown here appear to be more cumbersome when compared to simply writing
it out in other programming languages, this only holds when trying to define the
model using different types of transitions.  It can be seen above that an end
user can almost view it as writing the model as they would in MATLAB, by
replacing the equality sign with initialization of \code{Transition} object.

\begin{figure}[!h]
\centering
\includegraphics[width=5in]{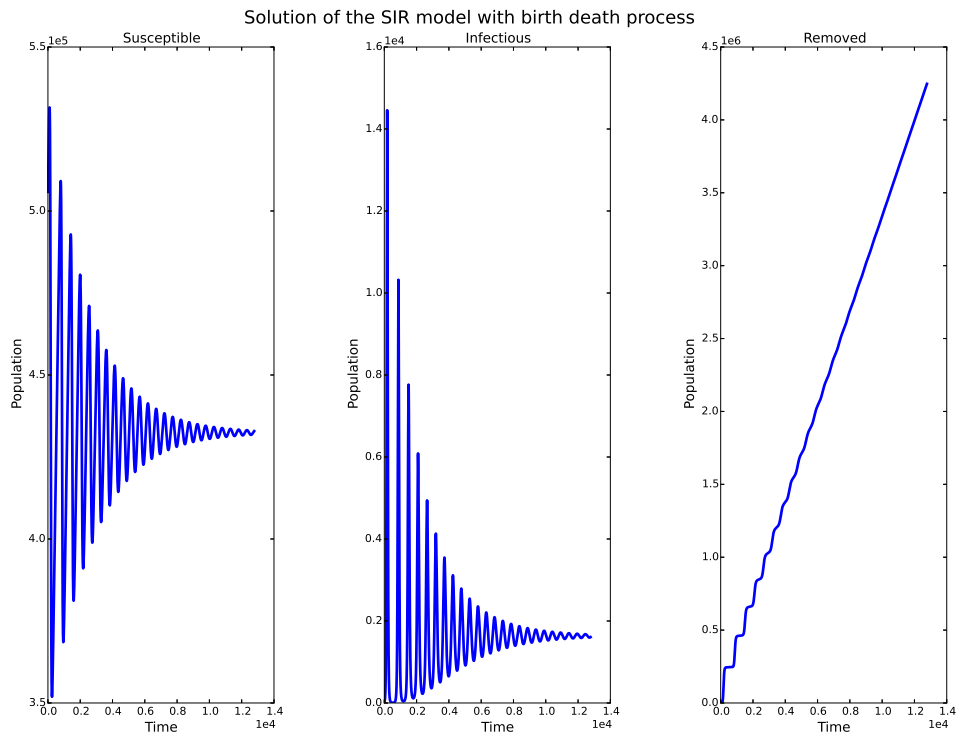} % {Fig3}
\caption{Solution of the SIR model over time with birth death processes
that induce oscillations}\label{fig:sir_bd_plot}
\end{figure}

All birth and death processes can be added to the model at any time, given that
the corresponding parameters exist in the model object.  Below, in the first
six commands, we add three birth/death processes to the original SIR
model, add the additional birth rate parameter and redefine the time--line.
These operations and setting the value of the new parameters can be done without
referring to information previously defined. The last line of the code
simply recomputes the solution given our new system, and the corresponding
plot in Fig. \ref{fig:sir_bd_plot}.
\begin{lstlisting}[language=Python] 
    >>> bdList = [Transition(origin='S', 
                             equation='B', 
                             transition_type=TransitionType.B),
                  Transition(origin='S', 
                             equation='mu*S', 
                             transition_type=TransitionType.D),
                  Transition(origin='I', 
                             equation='mu*I', 
                             transition_type=TransitionType.D)]
    >>> B = 126372.0/365.0
    >>> t = np.linspace(0,35*365,10001)
    >>> model.param_list = ['B', 'mu']
    >>> model.birth_death_list = bdList
    >>> model.parameters = {'B':B,'mu':B/N}
    >>> solution = model.integrate(t[1::])                                                                           
\end{lstlisting}

%         - types of transitions
%This is the same result set as the other models constructed. 
An important point to consider is how the information regarding the construction
of ODEs is provided to \code{DeterministicOde} at initialization. For ODEs the
transition list is provided to the \code{ode} argument, for transitions to
the \code{transition} argument and for birth and deaths it is to the
\code{birth\_death} argument. \code{DeterministicOde} will raise an error if an
incorrectly typed transition is presented to these arguments. PyGOM has been
constructed in this way to capture common errors in model specification and to
help ensure that transitions are defined carefully. An ODE system may be
constructed with a mixture of transition types so long as the transitions are
placed in the correct list.

%     [Stochasity]
\subsection*{Stochastic simulation}
There are situations when we are less interested in just a single
deterministic solution to a model with given parameters, but in a set of
possible realizations given the variation and uncertainty in many natural
systems. In such cases, we are interested in the stochastic behavior of a
model. There are two common ways to introduce stochasticity to a model
\begin{enumerate}
  \item Take parameter values as realizations from a random process.
  \item Drive changes between states using a probabilistic jump process.
\end{enumerate}
PyGOM is capable of generating realizations for either of these two scenarios. 
Moreover, the manner in which a model is defined changes very little from the
deterministic case already discussed.   If the library \code{dask} is installed,
PyGOM will automatically generate realizations in parallel.

%         - Stocatsic parmeter
\paragraph{Parameter stochasticity}
When we wish to use the first type of stochasticity our parameter values are
drawn from an underlying distribution. For the SIR model to be biologically
meaningful it is clear that both $\beta$ and $\gamma$ must
be non--negative, so it would seem natural to use the gamma distribution. Some
of the more commonly used distributions are provided within the \code{utilR}
sub--package, where we have used the \href{http://www.r-project.org/}{R
language} \cite{r_core_team_r2014} naming conventions for the distribution
names and input argument. Users are free to use functions from
\code{scipy}'s \code{stats} sub--module or any other arbitrary function that is
\emph{callable} with the number of realizations as the first input argument
followed by the distribution parameters.

\begin{figure}[!h]
\includegraphics[width=5in]{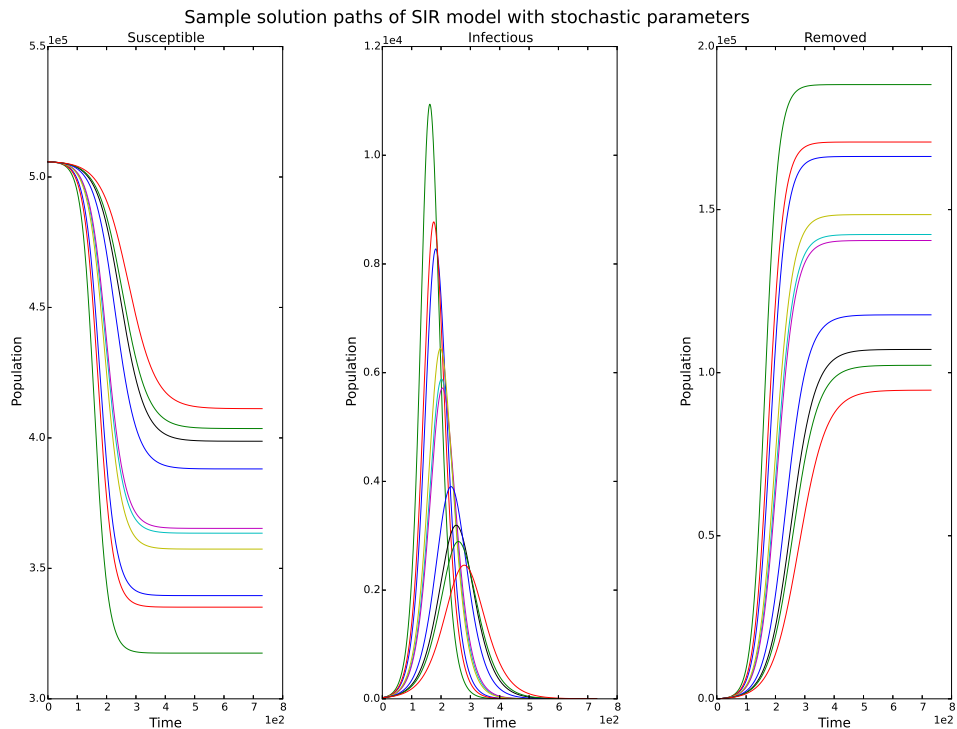} % {Fig4}
\caption{Results, by compartment, of 10 realizations of a
stochastic SIR model}\label{fig:stoch_plot}
\end{figure}

We define a stochastic model in a very similar way to the previous
models, indeed we can reuse the setup for the deterministic model defined above
\begin{lstlisting}[language=Python]
    >>> from pygom import SimulateOde
    >>> modelS = SimulateOde(states,
                             params,
                             transition=[sir_t1, sir_t2])
\end{lstlisting}
Now we define and set the parameters. We can use a mix of 
stochastic and non--stochastic parameters, if required, as shown below, where
the total population $N$ is a constant in this case and the birth and death
processes from above have been removed. Here we define the parameters in a
Python dictionary (\code{d}). Each parameter in the model is a \emph{key} in this dictionary
with the \emph{value} either as a constant or as a tuple containing the
generating function and a dictionary with the generating function's attributes
\begin{lstlisting}[language=Python]
    >>> from pygom.utilR import rgamma
    >>> d = dict()
    >>> d['beta'] = (rgamma, {'shape':3600.0, 'rate':1000.0})
    >>> d['gamma'] = (rgamma, {'shape':1000.0, 'rate':5000.0})
    >>> d['N'] = N
    >>> modelS.parameters = d
    >>> init_state = [int(i) for i in init_state]
    >>> modelS.initial_values = (init_state, t[0])
\end{lstlisting}
We generate 10 realizations (\code{iteration=10}) from this model as an example 
and ask for the full output of the simulations via \code{full\_output=True}
\begin{lstlisting}[language=Python]
    >>> Ymean, Yall = modelS.simulate_param(t, 
                                            iteration=10, 
                                            full_output=True)
\end{lstlisting}
The output from this simulation will be a tuple with the first
element containing the sample mean and the second a list of solutions. Here we
have simply split the tuple on assignment into \code{Ymean} and \code{Yall}.
The values in the \code{Yall} variable permit the user to construct an empiric
predictive interval and, by plotting the values in the \code{Yall} variable, we
may visualize the results of the simulation as in Fig. \ref{fig:stoch_plot}.

\begin{figure}[!h]
\centering
\includegraphics[width=5in]{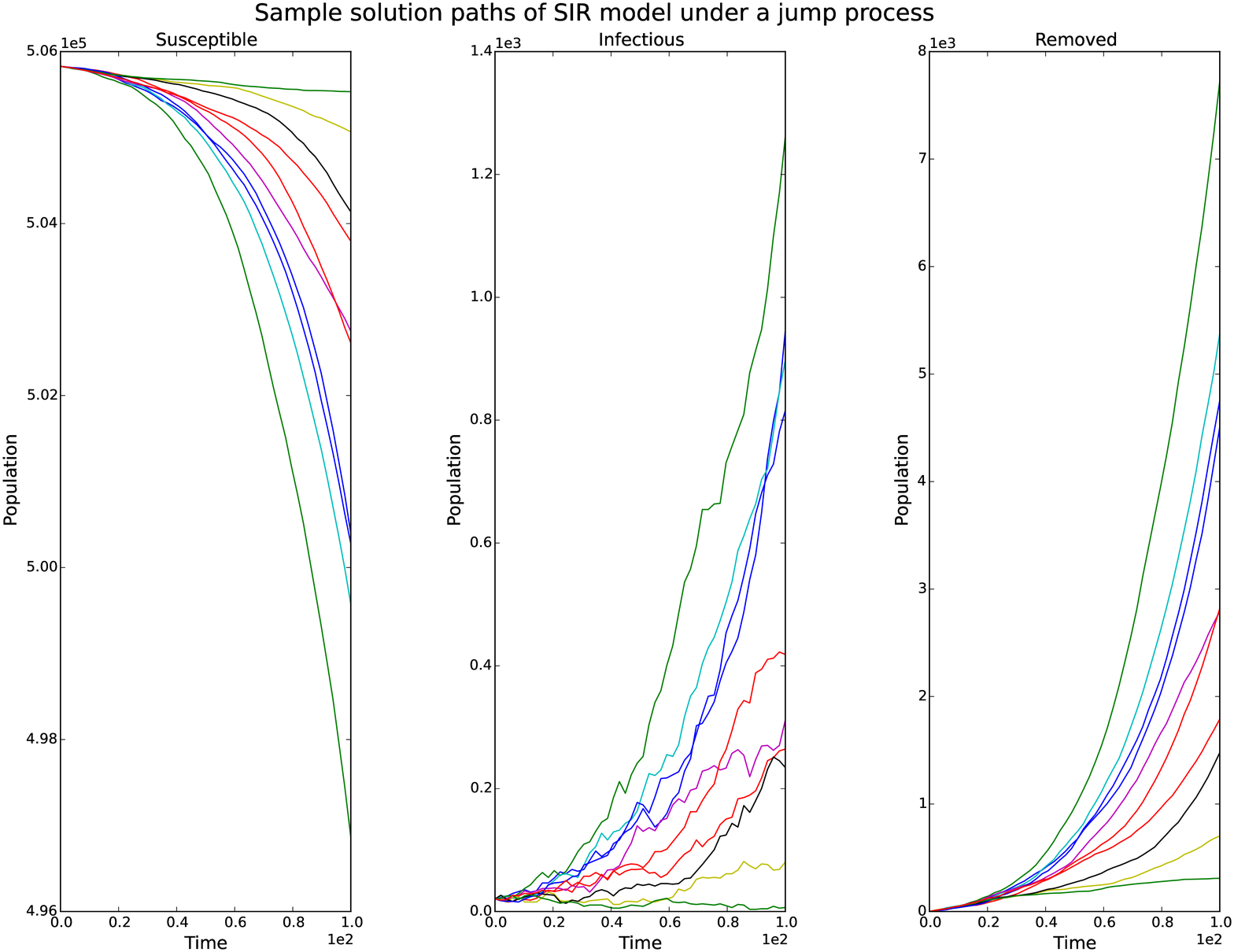} % {Fig5}
\caption{Ten simulated paths under a continuous time Markov
process.}\label{fig:sir_ctmc}
\end{figure}

%         - Stocastic transitions (Markov jump processes)
\paragraph{Jump processes or master equation stochasticity}
Compared to the example above where we assume that movements between states are
small and continuous, in this method of introducing stochasticity to
an ODE system we assume that movements between states are discrete, termed
jumps. More concretely, the probability of a move for transition $j$ is 
governed by an exponential distribution such that
\begin{equation}
\Pr(\textnormal{process $j$ jump within time } \tau) = \lambda_{j}
e^{-\lambda_{j} \tau}\nonumber
\end{equation}
where $\lambda_{j}$ is the rate of transition for process $j$ and $\tau$ 
the time elapsed after current time $t$. In chemistry and physics this known as 
a master equation model. Greater detail of these systems and their solutions
may be found in \cite{Gillespie2007}. We first reset the parameters so that they
are fixed rather than stochastic
\begin{lstlisting}[language=Python]  
    >>> modelS.parameters = param_eval
    >>> t_jump = np.linspace(0,100,50)
\end{lstlisting}
We then perform a set of jump process simulations, this is similar to parameter
stochasticity simulation, differing only in the name of the method
invoked
\begin{lstlisting}[language=Python]
    >>> simX, simT =modelS.simulate_jump(t_jump, 
                                         iteration=10, 
                                         full_output=True)
\end{lstlisting}
As before we can use the result variables with a graphics package to produce a
visualization of these simulations as in Fig. \ref{fig:sir_ctmc}.
Simulation results are approximate as they are performed using the
$\tau$--Leap algorithm \cite{Cao2006} by default, with the options of obtaining
an exact simulation \cite{Gillespie1977} if desired.

Here we have ``zoomed'' into a section of the time points compared to previous
Figures.  This is because the jumps occur on a much smaller time scale, and
indeed both the \textbf{S} and \textbf{R} state appears to be smooth with
discontinuity observed in only the \textbf{I} state.

Unlike any of the previous models a jump process model is able to produce
simulations where the disease is completely eliminated from the model before the
disease has run its full course (all members of the `\textbf{I}' compartment
moved to `\textbf{R}' before more individuals become infected). You can see the
result of this in Fig. \ref{fig:sir_ctmc} as the horizontal lines at the top
of the susceptible graph and at the bottom of the removed graph.

%     [Parameter estimation]
\subsection*{Parameter estimation and testing model fit}
Given an observational data set relating to a system being modeled
we may wish either to test an ODE based model to see how well it fits the data
or use the data to estimate the parameter values within the system. Were we to have
a set of $n$ observations $y_{i}$ at specific time points  $t_{i}$, $i =
1,\ldots,n$ we would require a function that measures the disparity between this
data and the model, a loss function. Within PyGOM we have implemented
the most common loss functions in the \code{ode\_loss} module. Of particular
note is the square loss (squared error) $\left\| \mathbf{y} - \hat{\mathbf{y}}
\right\|^{2}$ function which we use in the following examples.  Square loss is
also the simplest and most commonly used loss function. PyGOM also provides
parametric loss functions via the Poisson and Normal distributions.

All our loss functions come with the ability to return the cost, amount of loss
incurred with respect to the data, as well as the residuals which are essential 
to post--estimate analysis such as tests for normality and autocorrelation.
These loss classes take multivariate observations, \emph{i.e.} $\mathbf{y}$ is a
matrix of size $\left[ n \times k\right]$ where $n$ is the number of
observations and $k$ the number of targeted states. Furthermore, under the
square or normal loss functions, it is possible to set weights on the
observations. The weights may be scalar or vector, with size equal to the number
of targeted states or observations.

%estimating parameters
Parameter estimation is a non--linear optimization problem which
has been tackled by both deterministic and stochastic estimation methods
\cite{banga_global_2004}. That is, we seek a set of parameter values that
minimize the loss function.  Our focus here is on obtaining the derivatives
information as they are central in deterministic methods, and have been
shown to be useful in the stochastic setting such as Monte Carlo Markov
chain \cite{girolami_riemann_2011}.

We reuse the SIR model above, but this time initialized using the pre--defined 
version in the \code{common\_models} sub--module of PyGOM.  Note how easy it is
to use a completely different set of parameters, and with the corresponding
$x_{0}$ and $\mathbf{t}$, we are ready to solve the IVP.  The solution given
fixed parameters can be viewed as observational data with perfect information. 
Here, we scale the solution of the \textbf{R} states by a random multiplier to
ensure that the problem is non--trivial and take the result as our observed data
\begin{lstlisting}[language=Python]
    >>> from pygom import SquareLoss, common_models
    >>> model = common_models.SIR({'beta':0.5,'gamma':1.0/3.0})
    >>> init_state = [1,1.27e-6,0]
    >>> model.initial_values = (init_state, t_jump[0])
    >>> solution = model.integrate(t_jump[1:])
    >>> y = solution[1:, -1]
    >>> y = y * (0.90 + np.random.rand(len(y))/5.0)
\end{lstlisting}
Using this pseudo--data, the ODE object, our time and initial state vectors, we
now construct a square loss object with an initial guess for the parameters 
$\beta$ and $\gamma$, in \code{theta} below.
\begin{lstlisting}[language=Python]
    >>> theta = [0.5,0.5]
    >>> obj_sir = SquareLoss(theta=theta, 
                             ode=model, 
                             x0=init_state, 
                             t0=t_jump[0], 
                             t=t_jump[1:], 
                             y=y,
                             state_name=['R'])
\end{lstlisting}
In the example above we are looking at the entire parameter set (both  $\beta$
and $\gamma$) but only through values observed in the 'R' state.
However, it is perfectly possible to target only specific parameters instead of
the full set by specifying them through \code{target\_param} and to include other
state values through \code{state\_name}.

We are going to put some constraints on the parameter space where we think the
optimal parameter value may lie.  This is necessary for the SIR model because the
parameters must be non--negative, as per model definition. So,
we bound the value for both parameters to between $0$ and $2$.  These
bounds are specified in the same order as the parameters were constructed above.
\begin{lstlisting}[language=Python]
    >>> bounds = [(0.0, 2.0), (0.0, 2.0)]
\end{lstlisting}
In the following example we use the default
optimization method from \code{scipy.optimize},  with the gradient obtained from
forward sensitivity
\begin{lstlisting}[language=Python]
    >>> from scipy.optimize import minimize
    >>> theta_hat = minimize(fun=obj_sir.cost,
                             jac=obj_sir.sensitivity,
                             x0=[0.5, 0.5],
                             bounds=bounds)
\end{lstlisting}
In the result object the \code{x} gives the estimated parameter values. Here
the estimates were $\beta = 0.48427416, \gamma = 0.31797725$.

\begin{figure}[!h]
\centering
\includegraphics[width=5in]{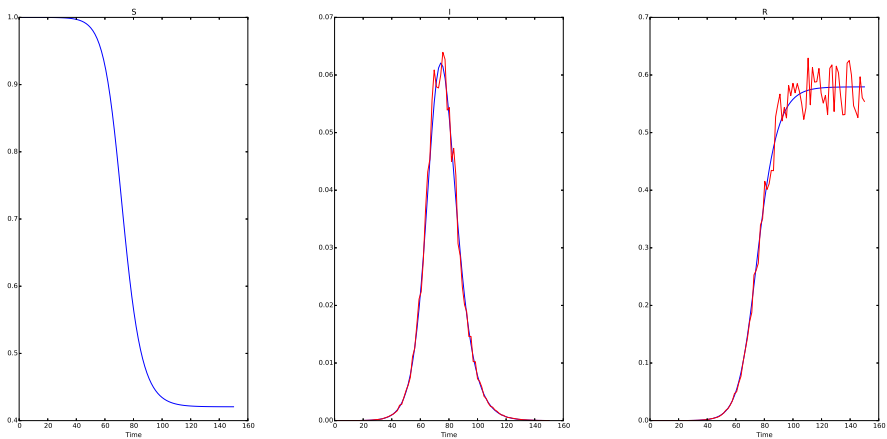} % {Fig6}
\caption{An example of the plot method from our loss classes.  The red line is
the observational data and the blue in all panels are the simulated paths with 
parameters as fitted.}\label{fig:sir_est}
\end{figure}

To visualize the \emph{goodness--of--fit} a plot method has been implemented
within the loss function class. This may be invoked by simply calling the
\code{plot()} method.   Fig. \ref{fig:sir_est} was generating using this 
convenience method which plots the observed values against the solutions
generated by the best--fit parameters
\begin{lstlisting}[language=Python]
    >>> obj_sir.plot()
\end{lstlisting}

\subsection*{Derivative Information}
As seen above we made use of the loss function's gradient when
estimating the unknown parameters. PyGom's loss functions
provide two ways to calculate this gradient: \code{sensitivity} and
\code{adjoint}, see 2.2 and 2.3 of \cite{Chavent2010} for details.
The \code{gradient} function by default is a synonym for \code{sensitivity}.
Substituting \code{adjoint} in place of \code{sensitivity} in the optimization
above only has impact on the computational speed, which
 depends on the properties of the ODEs and we refer interested readers to
 \cite{Li2000,Chavent2010}.
\begin{lstlisting}[language=Python]
    >>> S = obj_sir.sensitivity(theta)
    >>> A = obj_sir.adjoint(theta)
\end{lstlisting}

%\cite{Li2000,Chavent2010} references for forward sensitivity
%In testing the \emph{adjoint} method of calculating the gradient almost always 
%slower.  This is mainly due to the adjoint method requiring interpolation between 
%states which is relatively time consuming.

Additionally, Hessian information is also available via \code{hessian}.  The
Hessian for a non--linear problem is not guaranteed to be a positive
semi--definite matrix, hence certain algorithms such as the
Levenberg--Marquardt algorithm only uses the approximation of the Hessian
$\boldsymbol{H} \approx \boldsymbol{J}^{\top}\boldsymbol{J}$ where
$\boldsymbol{J}$ is the Jacobian. This is also available via 
\begin{lstlisting}[language=Python]
    >>> J = obj_sir.jac(theta)
\end{lstlisting}
Note that when $y$ is a multivariate observation, the return by
\code{residual} is a matrix and \code{jac} is a matrix with $np$ (number of
observations $\times$ number of parameters) columns.  If the approximation is
required instead of just the Jacobian, it can be obtained using
\begin{lstlisting}[language=Python]
>>> JTJ = obj_sir.jtj(theta)
\end{lstlisting}

%\subsection*{Parameter / Forward sensitivity}
%\red{add this section in part from 2.4 of documentation} not sure it fits!
%     [Confidence intervals]
\subsection*{Confidence Interval of Estimated Parameters}
After obtaining the \emph{best} fit value for a parameter, it is natural to
report both the point estimate  and the confidence level at a given $\alpha$
(the false positive or Type I error rate, typically $5\%$). Within PyGOM we
provide several methods to calculate such a confidence interval and describe
three in detail below.

%         - asymtotic
\paragraph{Asymptotic}
The simplest method of calculating a confidence interval is to invoke the
normality argument and use the Fisher information of the likelihood 
\cite{Casella2001}. From the  Cram\'{e}r--Rao inequality we know that
\begin{equation*}
\textnormal{Var}(\hat{\theta}) \ge \frac{1}{I(\theta)}
\end{equation*}
where $I(\theta)$ is the Fisher information, which we take as the Hessian.  
The normality comes from invoking the central limit theorem. Obtaining an
estimate of this confidence interval with PyGOM is as simple as defining our
significance level $\alpha$, calculating our fit and determining the interval.
\begin{lstlisting}[language=Python]
 >>> from pygom import confidence_interval as ci 
 >>> alpha = 0.05 
 >>> xLower, xUpper = ci.asymptotic(obj=obj_sir, 
                                    alpha=alpha,
                                    theta=theta_hat['x'])
    >>> print(xLower)
[ 0.21941127  0.07131115]
    >>> print(xUpper)
[ 0.74913705  0.56464335]
\end{lstlisting}
The \code{xLower} and \code{xUpper} objects now contain the lower and upper
bounds for the parameters. As before with the fits, the parameter order is the
same as was specified when the model was created.

%         - Profile likelyhood
%         - Geometric profile likelyhood
\paragraph{Profile and Geometric likelihood}
Another approach to calculating the confidence intervals is to take each
parameter individually, treating the remaining parameters as nuisance variables,
hence the term \emph{profile}.  We provide a function within the
\code{confidence\_interval} module to obtain such an estimate, \code{profile}.
The solving of the system of equations for profile likelihood requires
only Newton like steps, possibly with correction terms as per \cite{Venzon1988}.
However, this is usually hard or even impossible for ODE systems
because the likelihood is not monotonic either side of the central parameter
estimate.  This is typically caused by a lack of observations, and is therefore
not an issue which an end user is able to address. In the face of this we
provide an alternative way to generate a result similar to profile
likelihood using the geometric structure of the likelihood. We follow the method
in \cite{Suresh1987}, which involves solving a set of differential equations. 
The confidence interval is obtained by solving an IVP from $t = 0$ to $1$ and 
is all handled internally via the \code{geometric()} function in PyGOM's 
\code{confidence\_interval} module. A more in--depth exposition of these types 
of likelihood estimation is provided in the supplementary material in S4. %\ref{sec:profandgeom}
\begin{lstlisting}[language=Python]
    >>> xLGeometric, xUGeometric = ci.geometric(
                         obj=obj_sir,
                         alpha=alpha,
                         theta=theta_hat['x'])
    >>> print(xLGeometric)
[ 0.21371156  0.05306822]
    >>> print(xUGeometric)
[ 0.97617977  0.77965589]
\end{lstlisting}

%         - Bootstrap
\paragraph{Bootstrap confidence intervals}
Bootstrap estimation \cite{Davison1997} is a widely favored technique for
estimating confidence intervals. There exist many implementations of bootstrap.
A semi--parametric method seems to be the most logical choice within the context
 of ODEs (even when the assumptions are violated at times). When we say
 semi-parametric, we mean the exchange of errors between the observations.  Let
 our raw error be
\begin{equation*}
  \varepsilon_{i} = y_{i} - \hat{y}_{i}
\end{equation*}
where $\hat{y}_{i}$ will be the prediction under $\hat{\boldsymbol{\theta}}$
in our model.  Then we construct a new set of observations via
\begin{equation*}
  y_{i}^{\ast} = \hat{y}_{i} + \varepsilon^{\ast}, \quad
  \varepsilon^{\ast} \sim \mathcal{F},
\end{equation*}
with $\mathcal{F}$ being the empirical distribution of the raw errors. As with
the previous confidence interval methods \code{bootstrap} from the
\code{confidence\_interval} module will calculate this type of confidence
interval.
\begin{lstlisting}[language=Python]
    >>> xLBootstrap, xUBootstrap = ci.bootstrap(obj=obj_sir,
                                 alpha=alpha,
                                 theta=theta_hat['x'],
                                 iteration=100,
                                 lb=bounds_arr[:,0],
                                 ub=bounds_arr[:,1])
    >>> print(xLBootstrap)
[ 0.47400253  0.30820995]
    >>> print(xUBootstrap)
[ 0.49925899  0.33143419]
\end{lstlisting}

The bounds should be specified whenever possible because they are used when
estimating the parameters for all of the bootstrap samples.  An error will be
returned and terminate the whole process whenever the estimation process is not
successful.  All the bootstrap estimates can be obtained by setting 
\code{full\_output=True}, and they  can be used to compute the bias, tail 
effects and for tests of the normality assumption.  If desired, a simultaneous
confidence interval can also be approximated empirically.  Note however that 
because we are using a semi--parametric method here, if the model specification 
is wrong then the resulting information is also wrong.  The confidence interval
will still have the normal approximation guarantee if the number of samples is
large. As bootstrap confidence intervals follow an empiric distribution we do
not expect them to match those produced by the parametric types. In this case,
because the error in the observation is extremely small we find that the
confidence interval is narrower.

\section*{The PyGOM package}

\subsection*{Availability and Installation}
The source code for this package is available on 
\href{https://github.com/PublicHealthEngland/pygom}{GitHub} and through the
Python Package Index (PyPI). As such, it may be easily installed via pip:
\code{pip install pygom}

\subsection*{Dependencies}
Our package depends on most of the core SciPy libraries.  This includes SciPy,
Numpy, Matplotlib and Sympy.  Additional dependencies may be required,
which depends on the level of functionality the end user desires. For example, 
parallel stochastic simulation occurs if dask has been installed,
and displaying transition diagrams requires graphviz.

\section*{Conclusion}
To summarize, PyGOM is designed to simplify the construction of ODE
based models, with a bias towards modeling epidemiology that can: decomposing a
set of ODEs, obtaining epidemic related measures such as R0, perform parameter
estimation with high quality confidence intervals. This package is free and
will always be free.  Generic operations found for all types of ODE modeling are
available and plans to appeal to a wider audience by say integrating with the
SBML specification are underway.

Our intention was to make the definition of ODE based model systems easy while
maintaining rigor within that definition. This was to allow the rapid
assessment of published model systems and results in the absence of source code.
PyGOM has grown beyond that original idea into a comprehensive toolbox which
includes tools that make common operations on such systems, such as solving for
a series of time--points or fitting parameters to data, simple. In addition to
this software, within the \code{common\_models} module we have collected and 
implemented many common reference ODE systems as a foundation for the user to 
construct new models or fit canonical models to new data. To aid the newcomer 
to PyGOM, in addition to this paper included with the package there is an
extensive manual for PyGOM with further worked examples.

From the outset, the PyGOM package has been designed to be modular and
extensible. Starting with a small core of useful abilities, this modular
architecture has allowed new functionality to be added to the package without
requiring adjustments to existing code. Its modularity also helps keep the
code maintainable and comprehensible. Planned enhancements to the package will
seek to account for non--identifiability when parameter values are being
estimated and further assistance in the algebraic analysis of the ODE system.

As a system published under an Open License with the code freely available to
all, PyGOM fits well into the ever--expanding universe of Open Source analysis
tools. This openness permits the data--scientist to use PyGOM in conjunction
with other analysis libraries within Python as well as more widely with
other open--source tools such as those in R; and in environments ranging from
single machines through to large clusters and machine clouds.

By making common operations easy for the end user, we free them to use the
knowledge of their domain to construct and explore ODE--based model systems
without needing complex or esoteric computer code, leaving to the computer
the tedious tasks of book keeping and mathematical transformation. Even amongst
professional mathematicians and modelers PyGOM greatly simplifies and speeds up
the modeling process as it provides tools that allow easy construction of
robust work--flows, with validation and visualization aids built in. Together
these features allow the user to concentrate on the model.

\section*{Acknowledgments}
This work was supported by: the European Commission in the 7th Framework
Programme, (SEC--2013.4.1--4: Development of decision support tools for
improving preparedness and response of Health Services involved in emergency
situations) under grant number FP7--SEC--2013--608078 --- IMproving Preparedness
and Response of HEalth Services in major criseS (IMPRESS), the National
Institute for Health Research (NIHR) Health Protection Research Unit (HPRU) in
Modelling Methodology (NIHR--HPRU--2012--100--80) at Imperial College London,
the National Institute for Health Research (NIHR) Health Protection Research
Unit (HPRU) in Emergency Preparedness and Response (HPRU--2012--10414) at King's
College London in partnership with Public Health England
(PHE)

The views expressed are those of the authors and not necessarily those of
the NHS, the NIHR, the Department of Health or Public Health England.

\bibliography{ref}

\bibliographystyle{plain}

\end{document}